\documentclass[10pt,journal,compsoc]{IEEEtran}

\ifCLASSOPTIONcompsoc
  \usepackage[nocompress]{cite}
\else
  \usepackage{cite}
\fi

\usepackage{amssymb}
\usepackage{amsmath}
\usepackage{color}
\usepackage{graphicx}
\usepackage{algorithm,algorithmic} 

\newcommand{\revs}[1]{\textcolor{black}{#1}}

\newtheorem{theorem}{Theorem}

\begin{document}

\title{Policing 802.11 MAC Misbehaviours}

\author{Paul Patras\IEEEauthorrefmark{1},~\IEEEmembership{Member,~IEEE,} Hessan Feghhi\IEEEauthorrefmark{1}, David Malone,\\ and Douglas J. Leith,~\IEEEmembership{Senior Member,~IEEE}
\IEEEcompsocitemizethanks{\IEEEcompsocthanksitem P. Patras is with the School of Informatics, University of Edinburgh. 
\IEEEcompsocthanksitem H.~Feghhi, D. Malone and D. J. Leith are with the Hamilton Institute, National University of Ireland Maynooth.
\IEEEcompsocthanksitem (\IEEEauthorrefmark{1}) Joint first authors.
\IEEEcompsocthanksitem P. Patras was at the Hamilton Institute, National University of Ireland Maynooth when this research was conducted.
\IEEEcompsocthanksitem Work supported by Science Foundation Ireland grant 08/SRC/I1403.}
}


\IEEEtitleabstractindextext{
\begin{abstract}
With the increasing availability of flexible wireless 802.11 devices, the potential exists for users to selfishly manipulate their channel access parameters and gain a performance advantage.    Such practices can have a severe negative impact on  compliant stations. To enable access points to counteract these selfish behaviours and preserve fairness in wireless networks, in this paper we propose a policing mechanism that drives misbehaving users into compliant operation without requiring any cooperation from clients.   This approach is demonstrably effective against a broad class of misbehaviours, soundly-based, i.e. provably hard to circumvent and amenable to practical implementation on existing commodity hardware. 
\end{abstract}
}
\maketitle

\section{Introduction}
\begin{figure}[t]
 \centering\includegraphics[height=0.78\columnwidth, angle=270]{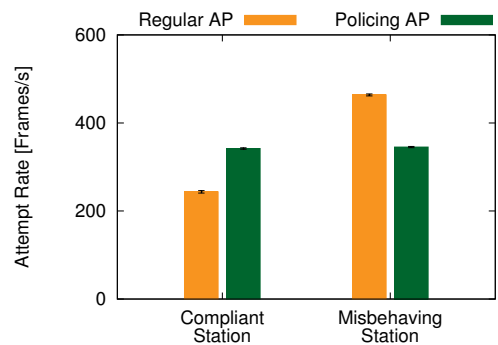}
 \caption{Wireless network with two stations, one contending with CW$_\text{min} = 32$ (compliant) and one with CW$_\text{min} = 16$ (misbehaving). Stations always have 1,000-byte packets to send and employ the IEEE 802.11 HR/DSSS physical layer at 11Mb/s. Average and 95\% confidence interval of the attempt rate attained by each station when the network operates with a regular AP, as well as with an AP running the policing scheme proposed in this paper. Experimental Data.}
 \label{fig:example}
\end{figure}

\IEEEPARstart{C}{omputers} equipped with WiFi devices that follow the popular IEEE 802.11 specification \cite{80211} employ a decentralised Medium Access Control (MAC) protocol to coordinate their transmissions on the channel. By design, this mechanism ensures compliant users connecting to a wireless network receive equal opportunities of accessing the medium and in this sense share resources in a fair manner. Each client station, however, operates independently and thus could act more aggressively in order to gain performance benefits, if changes can be made to the protocol behaviour. This already occurs in practice when network interface cards are not designed correctly, as reported in \cite{bianchi07}.  More critically, it can happen when users selfishly manipulate their channel access parameters to gain a performance advantage (see e.g. \cite{tang11}). This can cause significant unfairness, with the performance of the users that obey the standard being severely degraded \cite{raya06,liu08}. For example, 
consider a real network with two backlogged stations, one of them compliant and the other using a minimum contention window minimum (CW$_\text{min}$) half that recommended by the 802.11 standard. If the network operates with a regular access point (AP), the misbehaving user will transmit on average nearly twice as many frames as the compliant station. We illustrate this scenario in Fig.~\ref{fig:example} with light bars. Also plotted with dark bars is the performance of each client when the AP runs the policing scheme introduced in this paper, demonstrating its effectiveness in penalising misbehaving clients and equalising attempt rates, thereby restoring fairness.

Such MAC misbehaviours are increasingly of concern as open-source device drivers (e.g. MadWifi \cite{madwifi}, compat-wireless \cite{compat-wireless}, etc.) are becoming prevalent and permit users to modify the protocol rules either from the command line or with basic programming knowledge. Looking ahead, the trend is towards introducing still further flexibility, such as versatile architectures that allow changing the MAC operation of commodity hardware by reprogramming the protocol state machine with the help of simple visual tools~\cite{bianchi12}. 

In this paper we introduce an AP-based policing scheme for 802.11 Wireless LANs that is \emph{(i)} demonstrably effective against a broad class of misbehaviours, \emph{(ii)} soundly-based, i.e. provably hard to circumvent and, importantly, \emph{(iii)} amenable to practical implementation on existing commodity hardware. 
With this policing scheme, the AP controls the transmission attempt rate of misbehaving stations by acknowledging their frames with a probability that depends on the deviation of the stations' transmission attempt rate from the fair value. Decreasing the probability of acknowledgement causes a client station to backoff its contention window, thereby reducing its transmit rate and restoring fairness. An important feature of this approach is that it only requires measuring the transmit rate of each client station, which is straightforward as all traffic passes through the AP in the infrastructure operational mode, and does not require identification of the specific type of misbehaviour being performed (e.g. shorter backoff, frame bursting, etc.).

We provide a mathematical analysis of the proposed policing algorithm's convergence properties and prove its robustness in the presence of users that can detect APs that penalise misbehaviour. More precisely, we show that any strategy that seeks to game our policing algorithm, deviating from the fair operation, necessarily leads to lesser goodput performance for a misbehaving station in the long~run. 

To establish the feasibility of our proposal, we present a prototype implementation of the policing algorithm on off-the-shelf hardware.  We validate the performance of our implementation by conducting extensive experiments over a wide range of misbehaviour scenarios. The results obtained demonstrate that our solution effectively penalises misbehaviour irrespective of the network size, number of selfish users and the parameters manipulated, without impacting negatively the operation of compliant stations. We also show that our algorithm does not mistakenly penalise  compliant stations, even in complex situations where compliant stations generate different volumes of traffic and so some clients consume the air time underutilised by others. Further, we show that our proposal not only tackles MAC misbehaviour, but has no negative impact on state-of-the-art PHY rate control algorithms, while it successfully alleviates fairness issues that arise in practical deployments due to PHY/MAC interactions.

\revs{To the best of our knowledge, our proposal is the first AP-based MAC misbehaviour \emph{counteracting} solution with theoretical performance guarantees and a fully functioning prototype implementation that has been extensively evaluated by way of experiments conducted in a real Wi-Fi network. We summarise the key contributions of our work below.}
\revs{\begin{enumerate}
 \item[(1)] We design a novel algorithm that, unlike previous proposals, does not only address MAC misbehaviour detection, but thwarts selfishness without requiring non-trivial modifications of the protocol stack;
 \item[(2)] We specify a scheme that controls stations' transmission attempt rates and is robust to adaptive misbehaving strategies that seek to game its operation;
 \item[(3)] We provide detailed proof of this robustness and rigorous analytical evidence of the algorithm's convergence;
 \item[(4)] We detail a functional implementation of the designed system on real 802.11 hardware;
 \item[(5)] We give a sound methodology for estimating the maximum achievable attempt rate, without injecting traffic in the network or requiring changes to compliant stations;
 \item[(6)] We further validate the algorithm's convergence properties with real experiments;  
 \item[(7)] We provide a comprehensive performance evaluation of our scheme, running on commodity devices in a real deployment, covering a broad range of circumstances.
\end{enumerate}}

The rest of the paper is organised as follows. In Sec.~\ref{sec:related} we review related work.  In Sec.~\ref{sec:algorithm} we present the proposed policing algorithm and in Sec.~\ref{sec:analysis} we analyse its convergence properties and its robustness to misbehaviour strategies that seek to game its operation. In Sec.~\ref{sec:implementation} we detail the prototype we have implemented on commodity hardware and in Sec.~\ref{sec:evaluation} we report the results of the experimental evaluation conducted under different network scenarios. In Sec.~\ref{sec:nonideal} we investigate the operation of our solution under more problematic channel effects. Finally, Sec.~\ref{sec:conclusions} concludes the paper.

\section{Related Work}\label{sec:related}
Misbehaviour detection has received much attention from the research community (see e.g. \cite{cardenas04,kyasanur05,raya06,toledo07,serrano10,szott11,tang11,cardenas09}). Existing work, however, largely focuses on how undesired behaviour can be achieved with current cards and on engineering solutions that assist the AP in identifying disobedient users, as well as the nature of their misbehaviour \cite{raya06,serrano10,cardenas09}. Only a limited number of proposals address counteracting greedy actions, and these suffer from significant practical drawbacks. For instance, \cite{cardenas04} requires a reputation management system to prevent MAC layer misbehaviour, while a cross-layer interaction is assumed in \cite{kyasanur05} to enable higher layers to restrict the traffic that non-compliant clients generate. 

In contrast to prior work, in this paper we introduce an effective policing scheme for 802.11 Wireless LANs (WLANs) that overcomes the above limitations, as it does not require modification of the protocol stack and is amenable to practical implementation. By design, a key benefit of our policing algorithm is that it does not require any information about the number of active stations or the nature of their misbehaviour.

The underlying principle behind our approach is to control the attempt rate of misbehaving clients by censoring the generation of MAC layer acknowledgements (ACKs). ACK skipping has been suggested as a means to allocate bandwidth for traffic prioritisation in a network of well-behaved stations \cite{vollero04,vollero05,banchs10}, but to the best of our knowledge has not been implemented to date with real devices as this fundamental operation is handled at the firmware level. 

The solution we propose leverages our previous design \cite{dangerfield11}, but differs in that here: \emph{(I)} we aim to control the transmission attempt rate instead of throughput, thus seeking to equalise stations' air time \cite{checco11}. By driving the channel access probabilities of all clients to the same value, regardless of the contention parameters they employ, we effectively preserve short-term fairness. \emph{(II)} We allow carrying forward penalties, thus also achieve long-term fairness. Finally, \emph{(III)} we guarantee that the mechanism cannot be gamed by greedy users that detect its operation.

\section{Policing Algorithm}
\label{sec:algorithm}

In this section we first explain the class of misbehaviours our proposal tackles and then we detail the operation of the policing algorithm. \revs{We consider WLANs with a single-AP (or, alternatively a group of co-operating APs) operating in infrastructure mode, i.e. all packets are transmitted through the AP, as this is the default and most widespread operational mode of today's Wi-Fi deployments.}

\subsection{Class of Misbehaviours}
Our focus is on 802.11 MAC protocol misbehaviours.
We do not consider lower layer PHY attacks, e.g. ACK jamming, or higher layer selfish behaviour, e.g. TCP acknowledgement manipulation or station association attacks. We also confine consideration to behaviours that seek to obtain performance benefits, rather than simply to disrupt the network operation through e.g. signal jamming \cite{thuente07}, or exploiting security vulnerabilities \cite{edney2004}.   

Our interest in this class of greedy MAC behaviours arises from the observation that they can be realised with currently available open-source drivers that allow manipulation of the MAC layer parameters (CW$_\text{min}$, CW$_\text{max}$, AIFS and TXOP \cite{80211}), sometimes simply by issuing a single command on the system console (see e.g. \texttt{iwpriv} for Atheros-based cards). 
\revs{Note that, despite the possibility of broadcasting precise EDCA configurations by means of beacon frames from the AP, selfish clients are free to ignore any of the contention parameter values assigned through this (advisory) mechanism and the prevalence of such open drivers provides them sufficient incentives to do so.\footnote{\revs{Consequently, earlier TXOP-based airtime allocation approaches (e.g. \cite{tan04,tinnirello05}) do not provide effective policing when stations are misbehaving.}}}
We assume WLANs implement an authentication mechanism such as WiFi Protected Access (WPA2) \cite{80211i}, that prevents short and repeatedly aggressive sessions facilitated by MAC address spoofing techniques. \revs{Note also that the IEEE 802.11i standard ensures replay protection through several mechanisms, of which the use of CCMP (Counter Mode Cipher Block Chaining Message Authentication Code Protocol, Counter Mode CBC-MAC Protocol) or TKIP (Temporal Key Integrity Protocol) procedures are particularly relevant to our scheme. Thus, a selfish user will be unable to impersonate fair clients and jeopardise their reputation.} Our work can be adapted also to open-access networks, by augmenting it with a signal-strength based MAC layer spoofing detector \cite{shen08} \revs{or a passive device fingerprinting tool \cite{neumann12}. The resilience of our proposal to more sophisticated security attacks can be further strengthened if used in combination with fine-grained PHY layer information \cite{xiong13}.} 
\vspace*{-0.2em}

\subsection{Controller Operation}
To tackle this class of misbehaviours,  we propose that the AP exploits the fundamental nature of the acknowledgements within the ARQ mechanism of 802.11. Specifically, we use the fact that stations will increase their contention window and re-attempt to deliver a frame that was not acknowledged before sending the next packet. By appropriately suppressing ACK generation for cheating users, the AP can therefore reduce their transmission rate and drive them to fair operation.  

\revs{We consider WLANs that operate in a commercial setting where the service provider seeks to monetize connectivity and thus a na\"ive solution that simply disassociates users with marginal, possibly accidental misbehaviour (see e.g. \cite{bianchi07}), would be operationally unacceptable. Instead, our goal is to effectively correct such behaviours.} 
\revs{It is possible though that a misbehaving station does not increase its contention window despite not receiving ACKs. For such blatantly and deliberately misbehaving stations}, it is not possible to use ACK suppression to drive the station to fair operation and instead the policing algorithm adapts to drop all ACKs and associated data packets, reducing the goodput of such misbehaving stations to zero and eventually disassociating it from the network.

The key to the performance of this algorithm is the manner in which we adjust the rate of ACK suppression $P_{NACK,i}(t)$ for user $i$ at each time step $t$ of its execution. Algorithm  \ref{alg:police} details the operation of the proposed approach.
 
\begin{algorithm}
\begin{algorithmic}
\STATE Initialise $t=0$, $p_i(t)=0$,  $P_{NACK,i}(0) =0$  for client station $i, \forall i$. 
\LOOP
\STATE Estimate the maximum \emph{fair} transmission attempt rate $\bar{x}(t)$, given the current network conditions;
\FOR {each associated client station $i$}
\STATE Measure transmission attempt rate $x_{i}(t)$ of the station;
\STATE Update the penalty:
\begin{align}
\label{eq:update}
p_i({t+1}) = \max\left(0, p_i({t}) + \alpha\left(\frac{x_i({t})}{\bar{x}(t)} - 1\right)\right),
\end{align}
where $0<\alpha<1$ is a parameter that determines the speed of reaction to deviations from the fair behaviour; 
\STATE $P_{NACK,i}(t+1) = \min\{p_i({t+1}), 1\}$;
\STATE $t\leftarrow t+1$;
\ENDFOR
\ENDLOOP
\end{algorithmic}
\caption{Determining the rate of ACK suppression.}\label{alg:police}
\end{algorithm}
\vspace*{-0.25em}

For each station, the algorithm works as follows. At each execution step $t$, it compares the measured station's transmission attempt rate $x_i(t)$ against the fair value $\bar{x}(t)$.  When the attempt rate\revs{\footnote{\revs{We use the term ``attempt rate'' to refer to the stationary probability that a station transmits a frame in a randomly chosen slot time. Note that this does not refer to the PHY layer bit rate achievable with various modulation and coding schemes (MCS).}}} is above the fair value, the rate of ACK suppression is increased, and vice-versa when the attempt rate is below the fair value. Thus at a fixed point we have $x_i({t})/\bar{x}(t) - 1 = 0$, i.e.  $x_i({t})/\bar{x}(t)=1$ and consequently the station's attempt rate is driven to the fair value.\footnote{Note that, to streamline notation, we will often drop the $i$ subscript from now on, provided there is no scope for confusion.}

The algorithm requires an estimate of the maximum fair transmission attempt rate. That is, the transmit rate that would be achieved by a client station employing the standard recommended 802.11 MAC configuration. In Sec.~\ref{sec:rate_estimation} we discuss in detail how to estimate this quantity and show that the AP can perform this operation on commodity hardware, without requiring the cooperation of compliant stations.

Since $P_{NACK,i}(t)$ is a probability, it can only take values in $[0,1]$. However, as we do not impose an upper bound on the update of $p_i(t)$, we allow the algorithm to carry forward and accumulate the penalty when $p_i(t)-P_{NACK,i}(t) > 0$ (i.e. for aggressive behaviour where $P_{NACK,i}$ reaches 1), until the greedy station reverts to compliant operation. Thus we prevent gaining long-term advantage over compliant stations (see Sec.~\ref{sec:robustness}).

\begin{figure}[t]
\begin{center}
\includegraphics[height=0.75\columnwidth, angle=270]{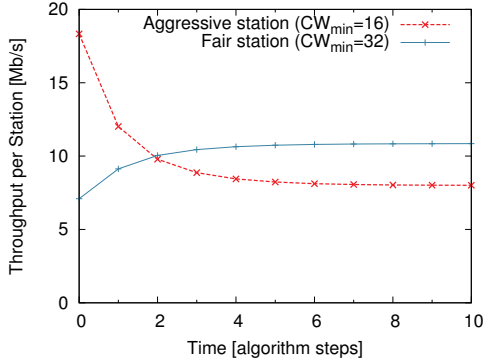}
\end{center}
\caption{Throughput performance in a Wireless LAN consisting of three saturated stations that transmit 1,500-byte packets using the 802.11 DSSS-OFDM physical layer at 54Mb/s. Two stations use the default MAC configuration (CW$_\text{min}$ = 32) and the third employs an aggressive setting (CW$_\text{min}$ = 16). The policing
algorithm is applied at the AP with $\alpha = 0.1$. Theoretical prediction.}
\label{fig:policingsim}
\end{figure}

Fig.~\ref{fig:policingsim} shows an example of the policing algorithm in operation.  
In this example we consider an 802.11g WLAN with three stations: two stations use standard contention
parameters and the third uses a smaller value of CW$_\text{min}$. Using a two-class Bianchi-like model \cite{malone07} we illustrate the time evolution of the stations' throughputs during the operation of the proposed policing scheme.
Observe that while the more aggressive
station initially claims more throughput due to the increased transmission attempt rate, the policing algorithm quickly adjusts the ACK drop probability, so that the misbehaving client receives lower performance. 

In what follows we provide a mathematical analysis of the the policing scheme's convergence and robustness properties and then present a practical implementation that we validate via extensive experiments in a real 802.11 Wireless LAN.

\section{Mathematical Analysis}
\label{sec:analysis}

In this section, we first establish the convergence properties of Algorithm~\ref{alg:police}. 
Second, we study the robustness of the proposed solution under misbehaviour strategies that seek to game its operation with the goal of achieving long-term performance benefits.

\begin{figure}[b]
\begin{center}
\includegraphics[height=0.77\columnwidth,angle=270]{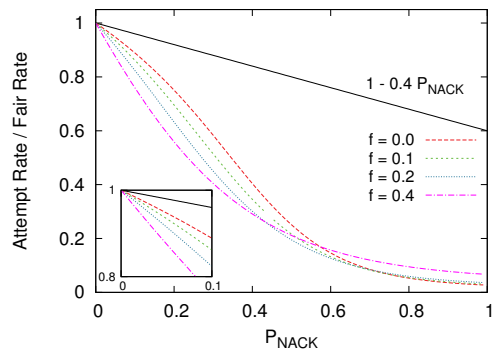}
\end{center}
\caption{The normalised attempt rate, $x(t)/\bar{x}(t)$, for a standard
compliant station over a range of network conditions (collision probabilities $f$) and ACK suppression rates $P_{NACK}$. The line $1 - 0.4 P_{NACK}$ represents an upper bound. Theoretical prediction.}
\label{fig:smoothtau}
\end{figure}

\subsection{Convergence}
We begin by establishing general conditions under which Algorithm \ref{alg:police}
converges to a fixed point. For well-behaved stations that follow the 802.11 distributed coordination function (DCF) specification, using a model such as \cite{bianchi00} we can verify that $\exists c$, $0<c<1$, such that $x(t)/\bar{x}(t) \le 1 - c P_{NACK}(t)$, $\forall t>0$. Specifically, the attempt rate of a fair station will
be proportional to the transmission probability, which we can calculate as a function of $P_{NACK}$, the failure probability $f$ seen by the station due to collisions, and other (fixed) MAC parameters. Fig.~\ref{fig:smoothtau}
shows that for a range of collision probabilities, these can be bounded with $c\le 0.4$.
Thus for well-behaved stations we have the following important result.

\begin{theorem}[Well-behaved stations]\label{thm:willconverge}
For stations satisfying $x(t)/\bar{x}(t) \le 1 - c P_{NACK}(t)$, $0 < c < 1, \forall t>0$, Algorithm~\ref{alg:police} ensures $\lim_{t\rightarrow \infty} p(t) = 0$.  That is, for well-behaved stations the policing algorithm does not drop any ACKs.
\end{theorem}
\begin{IEEEproof}
First note $p(t) \ge 0$ and if $p(t) = 0$, then subsequent terms $p(t+k)$, $k>0$, are
zero. If the sequence does not become constant at zero, then the $\max$
with zero is not active in Algorithm \ref{alg:police}, and we
consider two cases: 
\begin{enumerate}
 \item if $0 < p(t) \leq 1$, then
\begin{align*}
p(t+1) = p(t) + \alpha\left( \frac{x(t)}{\bar{x}(t)} - 1 \right)
	\leq p(t) - \alpha c p(t);
\end{align*}
 \item if $p(t) > 1$, then 
 \[
    p(t+1) \leq p(t) - \alpha c.
 \]
\end{enumerate}
So, at each step,
$p(t)$ decreases by at least $\alpha c \min(p(t),1)$. 	
Thus $p(t)$ is non-increasing and bounded below, and so convergent. As $p(t) - p(t+1) \rightarrow 0$, we see $\alpha c \min(p(t),1)\rightarrow 0$,
and thus $p(t) \rightarrow 0$.
\end{IEEEproof}
\vspace*{0.3em}

We  now show that in situations with misbehaving stations Algorithm \ref{alg:police}
also  converges.   Firstly, for misbehaving stations whose transmit attempt rates remain sensitive to ACK suppression, we have the following.
\begin{theorem}[Moderately misbehaving stations]
Suppose the transmit rate of a station satisfies the following conditions: 
\begin{enumerate}
 \item[i)] $x(t)/\bar{x}(t) > 1$ when $P_{NACK}(t) = 0$,
 \item[ii)]$x(t)/\bar{x}(t) < 1$ when $P_{NACK}(t)= 1$ and
 \item[iii)] $x(t)/\bar{x}(t)$ is strictly decreasing with $P_{NACK,t}$ and Lipschitz with a constant smaller that $2/\alpha$.
\end{enumerate}
Then Algorithm \ref{alg:police} converges to a point where $x(t) = \bar{x}(t)$.  
\end{theorem}

\begin{IEEEproof}
Since $x(t)/\bar{x}(t)$ is strictly decreasing, there exists a unique value
of $P_{NACK}(t)$ where $x(t)/\bar{x}(t) = 1$. We call this value $P$.  Let
$V(t) = \left(p(t) - P\right)^2$. Note that $V(t)$ is positive definite and radially
unbounded \cite{bacciotti06} in $p(t)$ and
\[
	V(t+1) = \left(p(t+1) - P\right)^2 \le
	\left(p(t) - P + \alpha \left( \frac{x(t)}{\bar{x}(t)} - 1 \right) \right)^2. 
\]
Expanding, we find
{\setlength\arraycolsep{0.1em}
\begin{eqnarray*}
	V(t+1) &\le& V(t) \\
		&+&\alpha
		\left(\frac{x(t)}{\bar{x}(t)} - 1\right)
		\left( p(t) - P \right)
		\left( 2 -
			\alpha \frac{\left(\frac{x(t)}{\bar{x}(t)} - 1\right)}
			     {p(t) - P}
		\right).
\end{eqnarray*}
}\\
Note that $\alpha > 0$ and $(x(t)/\bar{x}(t) -1)(p(t) - P)$ is strictly negative
except when $p(t) = P$, so if
\[
	2 > \alpha \frac{\left(\frac{x(t)}{\bar{x}(t)} - 1\right)}{p(t) - P},
\]
then we can ensure that $V(t)$ converges asymptotically to zero as
$t \rightarrow \infty$. However, this condition is ensured by requiring
$x(t)/\bar{x}(t)$ be Lipschitz in $P_{NACK}(t)$ (and consequently $p(t)$)
with a constant smaller that $2/\alpha$.  Thus, as $V(t) \rightarrow
0$ we have $p(t) \rightarrow P$.
\end{IEEEproof}
\vspace*{0.3em}

In the case of highly-aggressive stations for which the transmit attempt rate cannot be made fair using ACK suppression alone (e.g. when backoff of the MAC contention window has been disabled), we have the following.
\begin{theorem}
For stations where $\exists c>0$ such that $x(t) \ge \bar{x}(t)(1+c)$ for all $P_{NACK}\in [0,1]$, Algorithm \ref{alg:police} ensures $P_{NACK}(t) \rightarrow 1$. 
\end{theorem}
\begin{IEEEproof}
By assumption, $x(t)/\bar{x}(t) > 1$. Hence, $p(t+1) \ge p(t) + \alpha c$.
It follows that  $p(t)$ increases to a value greater than $1$ and
so $P_{NACK}(t) \rightarrow 1$. 
\end{IEEEproof}
\vspace*{0.3em}

Of course, some non-compliant stations may not meet the smoothness
conditions for convergence of $P_{NACK}$. Indeed, the station might
randomly choose an attempt rate at any time. However, in what follows we show
that in this case the station cannot gain from any such strategy.

\subsection{Robustness}
\label{sec:robustness}

Next we consider a scenario where a misbehaving client becomes aware of the policing algorithm running at the AP and attempts to game its operation, with the goal of achieving a long-term benefit in terms of throughput. We demonstrate that our scheme is robust to such sophisticated misbehaviour strategies by showing that, by design, the algorithm will penalise any strategy that deviates from the fair behaviour. 

Suppose that the selfish station seeks to maximise its goodput and remember
the algorithm can carry forward the penalty. The mean goodput
over the interval $[0,T]$ is given by
\begin{align}\label{eq:S}
S(T)&:=\frac{1}{T}\sum_{t=1}^T x(t)\left(1-p(t)\right) =\frac{\bar{x}}{T}\sum_{t=1}^T (1+y(t))(1-p(t)),
\end{align}
where $y(t)=x(t)/\bar{x}-1$. We can rewrite the policing update as
\begin{equation}
	p(t+1) = \max\left( 0, p(t) + \alpha y(t) \right), \label{eq:update2}
\end{equation}
and if we iterate this backwards to the previous time $t^*$ where
$p(t)$ was zero,\footnote{Note that $p(t)$ will be zero at least at $t^*=0$.} we see
\[
	p(t+1) = \max\left( 0, \alpha \sum_{k = t^*}^{t-1} y(k) \right).
\]
Suppose there is a time $T^* > 0$ with $p(T^*) = 0$ but $p(t) > 0$
for $1 \le t < T^*$. Then, we see $\sum_{k = 0}^{T^*-1} y(k) \le 0$,
so the average attempt rate of the station up to time $T^*$ is less
than that of a fair station. As $p({T^*}) = 0$, we may remove this
interval from our consideration and consider just the times from
$T^*$ onwards. By repeating this argument, we see that we only need
to consider the potential unfair behaviour of stations where $p(0) = 0$
and $p(t) = \alpha \sum_{k = 0}^{t-1} y(k) > 0$ for $1 \le t < T$.
We have the following result.
\begin{theorem}\label{th:one}
For policing Algorithm \ref{alg:police}, suppose that $\alpha \sum_{k
= 0}^{t-1} y(k) \ge 0$ for $1 \le t < T$. Let $Y$ be an upper bound
for $y(j)$ and let $\Delta > 1/\alpha + Y$ be a positive integer.
Then, if $T > \Delta$ and we consider the values of $S(T)$ as we
vary $y(1), \ldots, y({T-\Delta})$ and hold the other $y(j)$ fixed,
$S(T)$ is maximised by choosing $y(1) = \ldots = y({T-\Delta})
= 0$.
\begin{IEEEproof}
With policing update (\ref{eq:update2}) we have
\begin{align*}
p({t+1}) = \alpha\sum_{k=1}^{t} y(t),
\end{align*}
and we consider terms in $S(T)$ as follows.
\begin{align}
S(T)&=\bar{x} 
+\underbrace{\frac{\bar{x}}{T}\sum_{t=1}^T  y(t)}_{goodput\ gain}
-\underbrace{\frac{\bar{x}}{T}\sum_{t=1}^T\left(1+y(t)\right)p(t)}_{goodput\ cost}. \label{eq:S2}
\end{align}
Now,
\begin{eqnarray*}
\sum_{t=1}^T \left(1+y(t)\right) p(t) 
&&= \sum_{t=1}^T\left(1+y(t)\right)\alpha\sum_{k=1}^{t-1} y(t) \\
&&=\sum_{t=1}^T y(t) \alpha \sum_{k=t+1}^{T}\left(1+y(k)\right).
\end{eqnarray*}
So, the net relative gain is bounded by
\begin{eqnarray*}
&&\sum_{t=1}^T y(t) - \sum_{t=1}^T y(t) \alpha \sum_{k=t+1}^{T}\left(1+y(k)\right) \\
&& = \sum_{t=1}^T y(t) (1-\alpha(T-t)) - \alpha \sum_{t=1}^T \sum_{k=t+1}^{T} y(t) y(k). \\
\end{eqnarray*}
Taking the derivative with respect to $y(j)$ we get
\begin{eqnarray*}
\lefteqn{
(1-\alpha(T-j)) - \alpha \sum_{t \ne j} y(j)
} \\
&& = \alpha \left(\frac{1}{\alpha} - T + j - \sum_{t = j}^{T-1} y(t) + y(j) \right),
\end{eqnarray*}
which is negative when $j \le T-\Delta < T-1/\alpha - Y$, as the
sum is non-negative and $y(j) \le Y$. Thus, to maximise the gain,
we choose the smallest possible values of $y(j)$ subject to the
constraint on the partial sums being non-negative. Thus $y(1) =
\ldots = y({T-\Delta}) = 0$.
\end{IEEEproof}
\end{theorem}
\vspace*{0.3em}

This results confirms that no benefit can be obtained by deviating from the fair behaviour over $T-\Delta$ steps. Note however that a non-compliant client could potentially attempt to use a more aggressive transmit rate over the last $\Delta$ iterations before leaving the network, seeking to gain a small throughput benefit. But the fact that we allow for the penalty to carry forward to future times and consider networks that employ authentication makes such misbehaviours costly.

\section{Implementation}
\label{sec:implementation}

To demonstrate that deploying the policing algorithm is feasible with off-the-shelf hardware, in this section we present a Linux-based prototype implementation that we developed and discuss a non-intrusive technique for estimating the fair transmission attempt rate. 

\subsection{Prototype}
Implementing the suppression of MAC ACKs with existing devices is a challenging task, since generation of ACK frames is a basic operation that is handled at a low level within the wireless stack, below the device driver. To tackle this challenge, we based our implementation on an AP equipped with a Broadcom BCM4318 wireless adapter that employs the OpenFWWF firmware \cite{openfwwf}. The key advantage of using this open-source firmware is that it allows modifying the MAC protocol state machine running on the device, as already reported in \cite{han10,tinnirello12}. In addition to this, as the firmware runs on a modest 8 MHz processing unit on the network interface card, we modified the \texttt{b43} driver of the open-source \texttt{compat-wireless} package, to manage the more computationally demanding operations of our algorithm. 

\begin{figure}[b]
\begin{center}
\includegraphics[width=0.85\columnwidth]{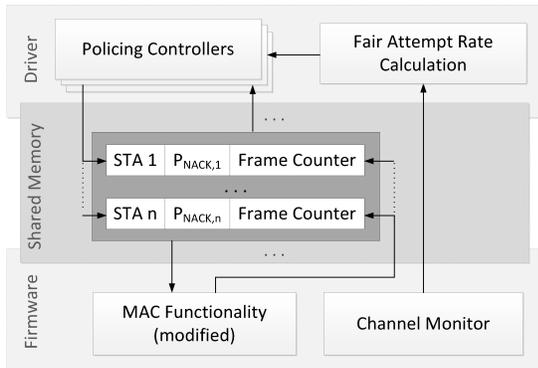}
\end{center}
\caption{Schematic view of the policing algorithm implementation. The policing update and fair rate estimation are implemented in the driver, per-station information is stored in the shared memory and ACK suppression is performed in firmware. }
\label{fig:schematic}
\end{figure}

Fig.~\ref{fig:schematic} illustrates the essential building blocks of our prototype. As shown in the figure, the implementation is split between the firmware and the driver: the former handles book keeping of per-station frame count, channel monitoring and ACK generation, while the latter manages the transmit rate computation and updating the ACK suppression rate for each associated client, based on the policing algorithm. To co-ordinate the operation of the firmware and driver modules, we rely on the 4KB \emph{shared memory}. We use this to store the information pertaining to each station and required by our algorithm, as we observe that a large portion of it remains unused during normal operation of the card. 

\begin{figure}[!t]
\begin{center}
\includegraphics[width=0.6\columnwidth]{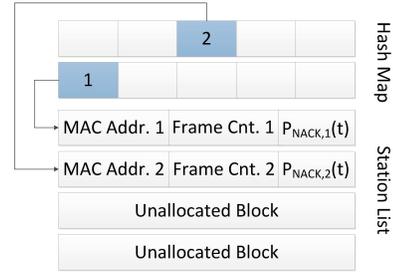}
\end{center}
\caption{Memory structure storing policing data. The hash map items point to per-station information elements storing the MAC address, frame counter (used to compute the attempt rate) and the current $P_{NACK}$ value.}
\label{fig:memorymap}
\end{figure}

We implement ACK handling in the firmware, as this is a highly time-sensitive operation. Specifically, the decisions to acknowledge or not a correctly received frame must be made within SIFS time and thus must not be interrupted or delayed by other tasks. For each frame received with a correct frame check sequence (FCS), we inspect the source MAC address, increment the frame counter (used by the driver to compute the attempt rate) of the sending station, fetch the corresponding $P_{NACK}$ value and decide to generate or suppress the acknowledgement. To complete these operations efficiently, our implementation employs a fast hash map and a list of information blocks. The hash-map consists of a 1 KB memory block that holds 512 2-byte pointers to sub-blocks storing the current frame count and ACK dropping probability associated to a station, as well as its MAC address. Fig.~\ref{fig:memorymap} shows the structure of the memory allocated for policing.

The policing update, which controls the penalty associated to each client, is implemented in the driver, as driver code runs on the CPU of the host and can perform calculations more quickly. The computation of the transmit rates and updates of the penalties according to (\ref{eq:update}) are executed at configurable discrete time intervals, when the driver reads the information stored in the shared memory for each associated station and performs the following operations: \emph{(i)} computes the transmission attempt rate of each station based on the frame count, \emph{(ii)} estimates the fair attempt rate (see Sec.~\ref{sec:rate_estimation}), \emph{(iii)}~updates the ACK dropping probabilities $P_{NACK,i}$ and writes their values back into the corresponding blocks, and \emph{(iv)} resets the frame counters. 

\subsection{Fair Attempt Rate Estimation}
\label{sec:rate_estimation}

To decide whether to police an associated station, our algorithm measures its performance and compares this to the maximum transmission attempt rate a fair client would attain under current network conditions. In this subsection, we discuss one mechanism for achieving fair attempt rate estimation non-intrusively, i.e. without injecting traffic in the network or requiring message passing between the AP and other stations. We will show that observing the wireless channel for a duration above 5 seconds ensures a good estimate of fair performance.

Towards this end we run a \emph{virtual MAC} instance at the AP, that reproduces the operation of a fair station, but does not release packets on the channel. Instead, we monitor channel slots and check the outcome of ``virtual'' transmissions, i.e. whether virtual attempts would have resulted in successes or collisions. Based on these observations, the mechanism estimates the failure probability $f$ experienced by a fair station, which can be then used to derive the attainable transmission attempt rate. In what follows, we give a formal analysis of this approach and investigate its accuracy.

Suppose we have a network of $n$ stations transmitting with probabilities
$x_1, \ldots, x_n$. Further, suppose that a station is saturated, for instance station 1. Assume for now that this station is fair.
We can write the failure probability due to collisions for this station as
\[
	f_1 = 1 - (1 - x_2) \ldots (1-x_n).
\]
As the station is fair,  
\[
	x_1 = g(f_1),
\]
where $g$ is a function mapping the failure probability to the transmission
probability and is given by \cite{wu02}:
\begin{eqnarray}
\label{eq:tau}
  g(f) &=& \frac{2(1-2f)(1-f^{R+1})} {W(1-(2f)^{m+1})(1-f)+(1-2f)(1-f^{R+1})}\nonumber \\
  & & \frac{~}{+W2^m f^{m+1}(1-2f)(1-f^{R-m})}.
\end{eqnarray}
In the above, we denote $W=$ CW$_\text{min}$, $m$ is the maximum backoff stage and $R$ denotes the retry limit. 

\begin{figure}[!t]
\begin{center}
\includegraphics[height=0.72\columnwidth,angle=270]{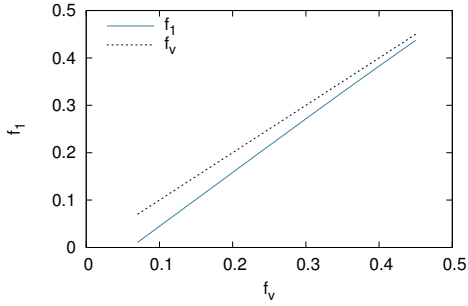}
\end{center}
\caption{Relationship between failure probability of a virtual station $f_v$ and that of a real fair client $f_1$. Theoretical prediction.}
\label{fig:pv-p1}
\end{figure}

Consider now that the AP runs a saturated virtual MAC instance. We can similarly express the failure probability $f_v$ this observes, as follows:
\begin{align*}
	f_v &= 1- (1-x_1)(1 - x_2) \ldots (1-x_n) \\
	     &= (1-x_1) (1-f_1) = 1- (1- g(f_1)) (1-f_1),
\end{align*}
where $g$ is the fair backoff function given by (\ref{eq:tau}). Note that if we know $f_v$,
we can solve the above for $f_1$. In Fig.~\ref{fig:pv-p1}, we plot the relationship between the virtual and actual failure probability of a saturated station. To add perspective, we also plot $f_v$ with a dotted line; we observe that the difference between the two is relatively small and reduces as the contention rate increases. 

Since there is a one-to-one mapping from $f_v$ to $f_1$, we can invert this to obtain an exact value for the failure probability of a fair saturated station and apply (\ref{eq:tau}) to compute the maximum achievable rate $\bar{x}$ of a fair station. 

The remaining question is how long should the channel observation period be, to ensure an accurate estimate of $f_v$. To answer this, we regard the virtual transmission attempt as a Bernoulli trial, whereby assuming independent trails, a failure is observed with probability $\hat{f_v}$ and a success with probability $1-\hat{f_v}$. By the central limit theorem, if the number of observations $N$ is large, the distribution of $\hat{f_v}$ is approximately normal with mean $f_v$ and variance $\sigma^2=f_v(1-f_v)/N$.

\begin{figure}[t]
\begin{center}
\includegraphics[height=0.72\columnwidth,angle=270]{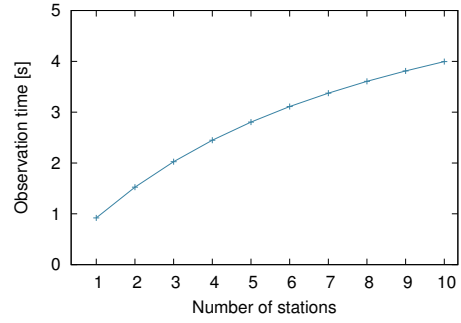}
\end{center}
\caption{Observation time required to estimate the collision probability $f_v$ of a fair client as the number of active station increases. Theoretical prediction.}
\label{fig:time_obs}
\end{figure}

\begin{figure*}[!t]
\normalsize
\centering
  \includegraphics[width=0.92\columnwidth,angle=270]{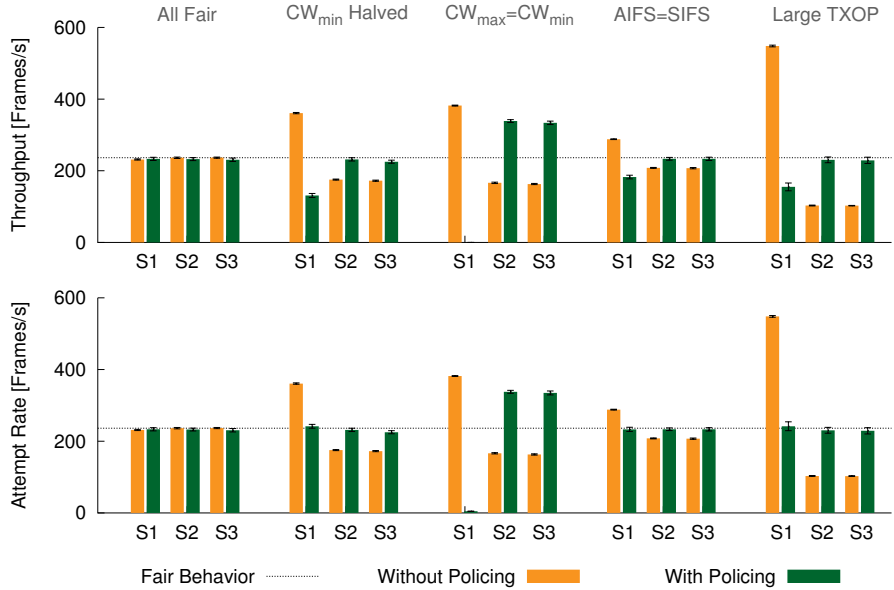}
  \caption{WLAN consisting of three backlogged stations sending 1,000-byte packets using the IEEE 802.11 HR/DSSS physical layer at 11Mb/s. Station S1 employs one of four types of MAC misbehaviour, stations S2 and S3 are standard compliant. Average throughput (above) and attempt rate (below) of each station in each scenario, when the network operates with a regular AP (light bars) and an AP running our policing algorithm (dark bars). Also plotted is the performance of a station when all clients are fair. Experimental data.}
  \label{fig:performance}
\end{figure*}

Say we want to compute the number of samples $N$ that gives us 95\% confidence that the estimated mean has precision $\epsilon$, i.e. $P(|f_v-\hat{f_v}| > \epsilon) < 0.05$. The confidence interval is $\hat{f_v}\pm z\sigma$, where $z=1.96$ is the z-score required for 95\% confidence. Since $\sigma$ is unknown and $\hat{f_v}(1-\hat{f_v}) \leq 0.5$, using this conservative upper bound \cite{shafer2012}, $N$ must satisfy
\begin{align*}
 \frac{z}{2\sqrt{N}} = \epsilon.
\end{align*}
Thus the number of observations required to ensure a good estimate of the fair attempt rate is
\begin{align*}
N = \left(\frac{z}{2\epsilon}\right)^2. 
\end{align*}
To translate this into an observation period required for a good estimate of fair performance before an update of the $P_{NACK}$ probabilities, consider the average slot duration in a network with saturated stations
\begin{align*}
E[T_{slot}] = P_e\sigma+P_sT_s+P_cT_c, 
\end{align*}
where $P_e$, $P_s$ and $P_c$ are the probabilities that a slot is empty, contains a success and respectively a collision, and $\sigma$, $T_s$ and $T_c$ are the corresponding slot durations (see \cite{bianchi00} for detailed calculations). Thus we compute the observation interval that gives an accurate estimation of the mean as\footnote{Note that $T[slot]$ is upper bounded by the length of a successful transmission $T_s$, which is readily obtainable in practice from the ``duration'' field of correctly received frames. Thus, one could avoid the complexity of computing $T_{slot}$ and use $T_s$ instead, to simplify implementation.} 
\[
T_{update} = N \cdot E[T_{slot}]. 
\]
To indicate the values $T_{update}$ would take in practice for $\epsilon=0.01$, in Fig.~\ref{fig:time_obs} we plot the necessary channel observation time for obtaining an estimate according to the above requirements for different network conditions in terms of number of saturated stations and assuming stations send packets with 1,000-byte payload at 11~Mb/s (IEEE 802.11 HR/DSSS). We conclude, that an observation interval above 5 seconds will ensure a good estimate of the fair performance in many scenarios. In our experiments we conservatively use a $T_{update} = 10$s for all tests.

In what follows, we evaluate the performance of our prototype in a real testbed and demonstrate its effectiveness under different types of misbehaviour.

\section{Experimental Evaluation}
\label{sec:evaluation}

Having described the design and implementation of our proposal, we now evaluate the performance of the policing algorithm in a real 802.11 testbed and prove its effectiveness under different types of misbehaviours and a wide range of network conditions. Our deployment consists of nine Soekris net4801 embedded PCs, one acting as AP and the other eight as stations. The AP is equipped with a Broadcom BCM4318 wireless card and is capable of running our prototype. The clients use Atheros AR5212 chipset adapters and the \texttt{ath5k} driver, which we modified to allow manipulating the MAC parameters by simple commands from the system console. All clients employ the 802.11 HR/DSSS physical layer (802.11b) and, if not otherwise specified, do not perform rate adaptation. 

Unless stated otherwise, we consider all stations are backlogged and send unidirectional UDP traffic to the AP. In all cases, we measure the performance of the stations when the network is operating with a standard AP and an AP running the proposed policing algorithm configured with the following settings: speed of reaction factor $\alpha=0.1$ (see (\ref{eq:update})) and update period $T_{update} = 10$s. 

\subsection{Controller Validation}
First we study the impact of four types of misbehaviour that can be easily implemented with current hardware, whereby aggressive MAC settings are used. Specifically, we investigate the scenarios where a user seeks to obtain performance benefits by employing selfish configurations as follows: \emph{(i)} contending with a CW$_\text{min}$ parameter half the default value (``CW$_{\text{min}}$ Halved''), \emph{(ii)} disabling the Binary Exponential Backoff (BEB) mechanisms while keeping a smaller CW$_\text{min}$ setting (``CW$_{\text{min}}$=CW$_{\text{max}}$''),\footnote{Note that compliant devices employ CW$_\text{max} > $ CW$_\text{min}$ settings to reduce failure probability upon subsequent attempts, thus being less aggressive.} \emph{(iii)} using a shorter interframe space post-backoff (``AIFS = SIFS''),\footnote{AIFS $\geq 2\sigma +$ SIFS is the amount of time a station is required to sense the channel idle before entering the backoff procedure. SIFS=10$\mu$s is the short interframe space. $\sigma$ is the 
duration of an idle slot.} and \emph{(iv)} retaining the access to the medium for 6.413ms by violating the TXOP$_{\text{limit}}$ parameter (``Large TXOP''), thus being able to send multiple frames upon a single transmission. 

In these scenarios we consider a simple network topology with one misbehaving stations sharing the medium with two fair clients that contend for the channel using the default MAC parameters specified by the 802.11 standard (i.e. CW$_\text{min}=32$, CW$_\text{max}=1024$, AIFS = DIFS = 50$\mu$s, TXOP = 0). Each client is saturated and transmits 1,000-byte packets to the access point for a total duration of 3 minutes. We measure the throughput and attempt rate of each station under each scenario, with and without the policing algorithm running at the AP, and repeating 10 times each test to compute average and 95\% confidence intervals with good statistical significance. 

Fig.~\ref{fig:performance} shows the throughput and attempt rate attained by each client in each of the scenarios considered, both with and without our policing algorithm running at the AP. To add perspective, we also plot with a dotted line the performance of one station when when all clients behave fairly (``All Fair'').
Observe that a selfish user using a smaller CW$_\text{min}$ attains nearly twice the throughput of compliant stations if not policed, whilst reducing the throughput and attempt rate of the fair users (``CW$_\text{min}$ Halved'', light bars). When we activate the policing algorithm (dark bars), this behaviour is effectively counteracted, as our solution equalises the attempt rates, while the misbehaving client sees its throughput performance reduced. If the selfish behaviour becomes more aggressive (``CW$_{\text{max}}$=CW$_{\text{min}}$'', light bars), e.g. the cheater employs a fixed contention window and thus does not backoff upon failures, in fact the policing algorithm rapidly increases the ACK dropping probability corresponding to that client to 1, thereby disassociating this from the AP. This is reflected in both the attempt rate and throughput performance, which are effectively zero when policing is applied (dark bars).

\begin{figure}[t]
\begin{center}
\includegraphics[height=0.73\columnwidth,angle=270]{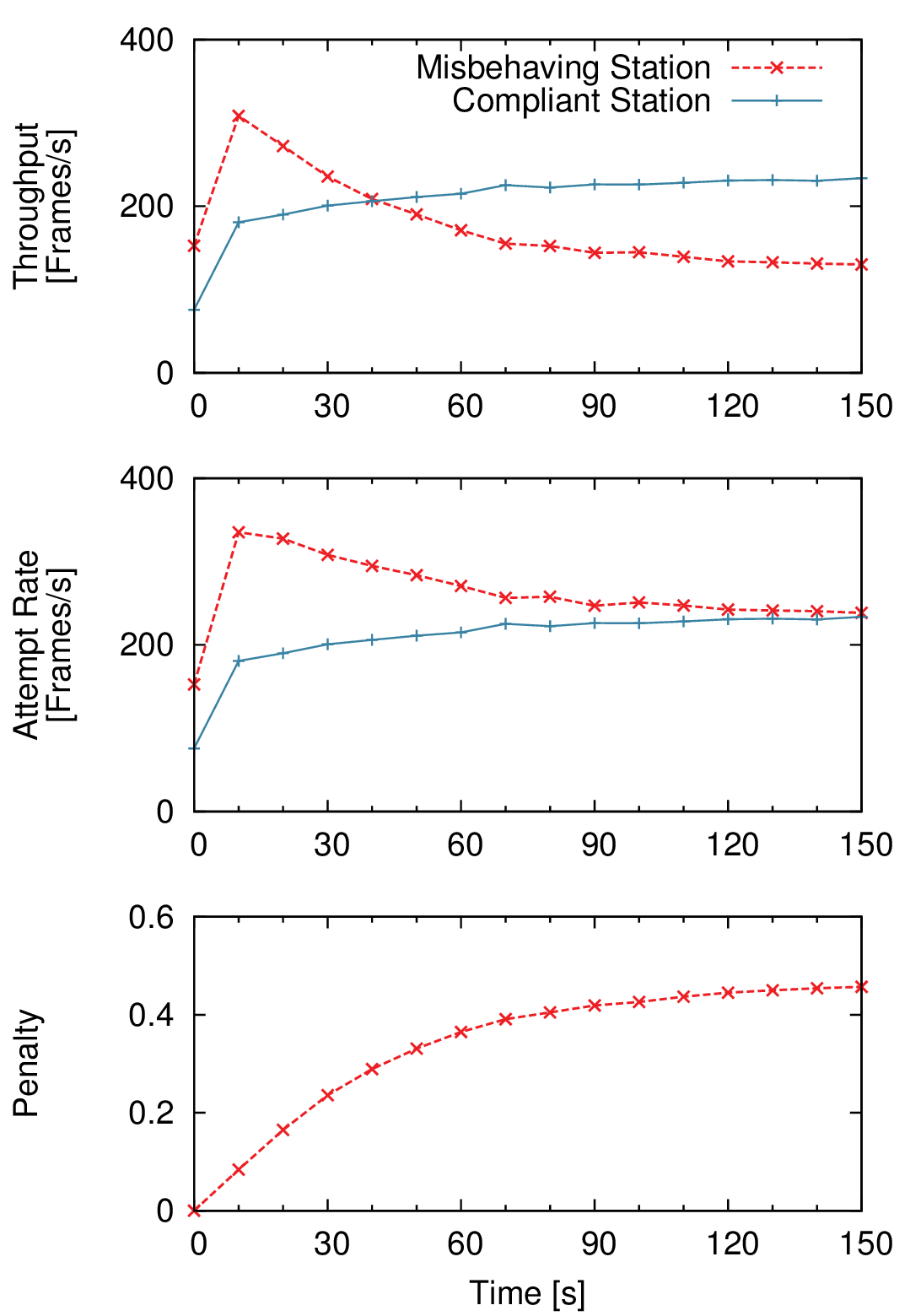}
\end{center}
\caption{WLAN consisting of three saturated stations: two compliant and one misbehaving, using a CW$_{\text{min}}$ half the default value. The AP runs the proposed policing scheme. Time evolution of the throughput (above), attempt rate (middle) and penalty applied by the proposed policing algorithm (below) for the misbehaving station and one fair client. Experimental data.}
\label{fig:details_cwmin}
\vspace*{-0.75em}
\end{figure}

\begin{figure}[t]
\begin{center}
\includegraphics[height=0.73\columnwidth,angle=270]{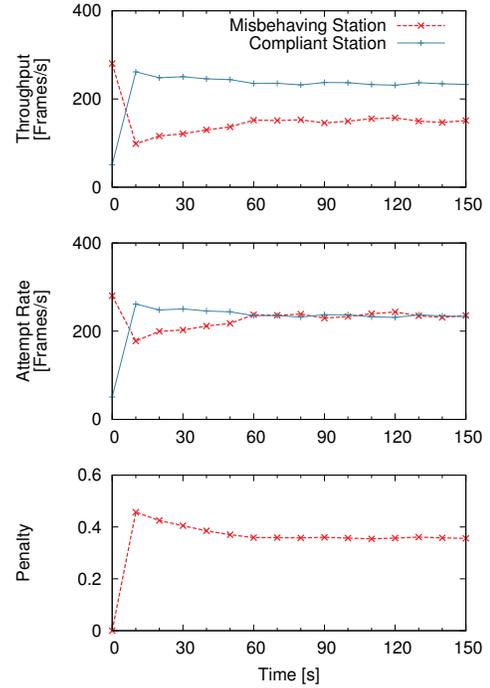}
\end{center}
\caption{WLAN consisting of three saturated stations: two compliant and one misbehaving, using TXOP = 6.413 ms. The AP runs the proposed policing scheme.
Time evolution of the throughput (above), attempt rate (middle) and penalty applied by the proposed policing algorithm (below) for the misbehaving station and one fair client. Experimental data.}
\label{fig:details_txop}
\vspace*{-0.5em}
\end{figure}

A more subtle misbehaviour strategy could employ a short post-backoff interframe space, e.g. the greedy station only waits SIFS before a new attempt, which is the minimum time separating two consecutive frames. Although less significant (since the selfish station sometimes randomly selects a large backoff counter and waits more than the other contenders that wait DIFS plus a short backoff value), the non-compliant client still achieves performance gains to the detriment of the fair stations present in the network (``AIFS=SIFS'', light bars). Once again, if we execute the policing algorithm at the AP, the transmission attempt rates are equalised and fairness is restored also in this case (dark bars).

Lastly, if the misbehaving user transmits several frames upon a single channel access (``Large TXOP''), their throughput performance is significantly higher than that of the fair stations as no action is taken to correct this selfish comportment (light bars). In contrast, with the proposed policing scheme, attempt rates stay equal and the cheater sees their throughput throttled down below the value corresponding to fair operation (dark bars).

Let us now take a closer look at the behaviour of the controller implemented by our scheme. Specifically, we are interested in validating the convergence of the algorithm under different types of misbehaviour. For this purpose, we pick two of the four scenarios discussed above and examine the time evolution of the network performance. More precisely, in Figs.~\ref{fig:details_cwmin} and~\ref{fig:details_txop} we show the time evolution of the
throughput and attempt rate for the non-compliant user and a fair station, as well as the penalty applied by our algorithm, in the cases when the selfish client uses a CW$_\text{min}$ half the default value and respectively a large TXOP setting, e.g. TXOP = 6.413ms.

In both cases, observe that the policing algorithm successfully brings the attempt rate of the misbehaving station down to that of a fair client (middle graph), while their throughput is reduced (top graph). What is important to remark is that the algorithm is close to convergence after a few steps, with the convergence time being shorter for more aggressive behaviour (i.e. with manipulated TXOP). Note also that the convergence time can be further reduced by choosing a larger $\alpha$ parameter.

\subsection{Impact of Network Size}
Next, we investigate whether a misbehaving client could hide in the crowd as the number of network users increases. For this purpose, we consider a network with one selfish station employing a small CW$_\text{min}$ based misbehaviour and we vary the number of fair stations, while we examine the performance of both. In each case, all clients are backlogged and send 1,000-byte packets for a total duration of 3 minutes. We repeat each experiment 10 times and compute the again average with 95\% confidence intervals of the attempt rate and throughput attained by each station.

\begin{figure}[t]
\begin{center}
\includegraphics[height=0.73\columnwidth,angle=270]{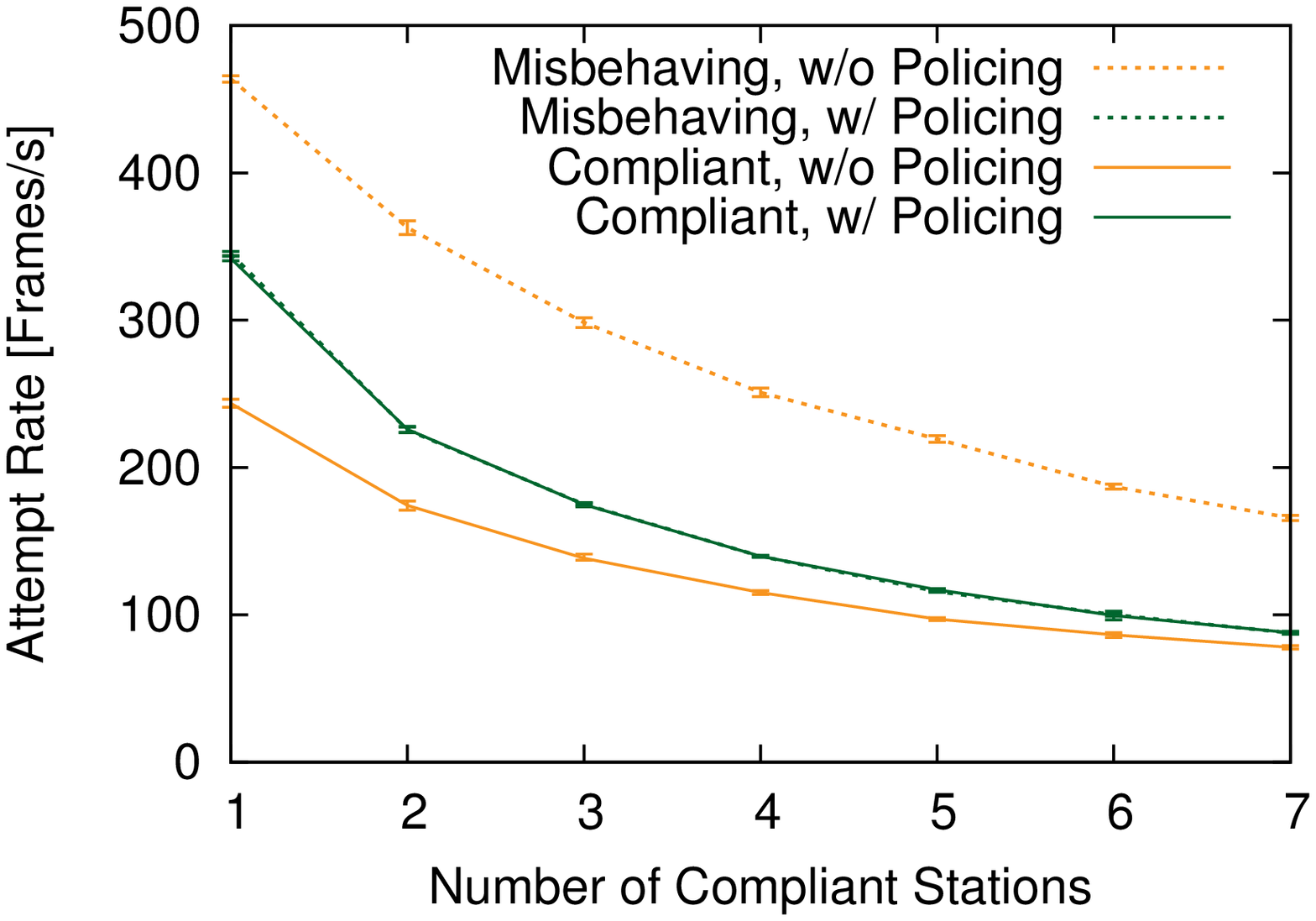}
\includegraphics[height=0.73\columnwidth,angle=270]{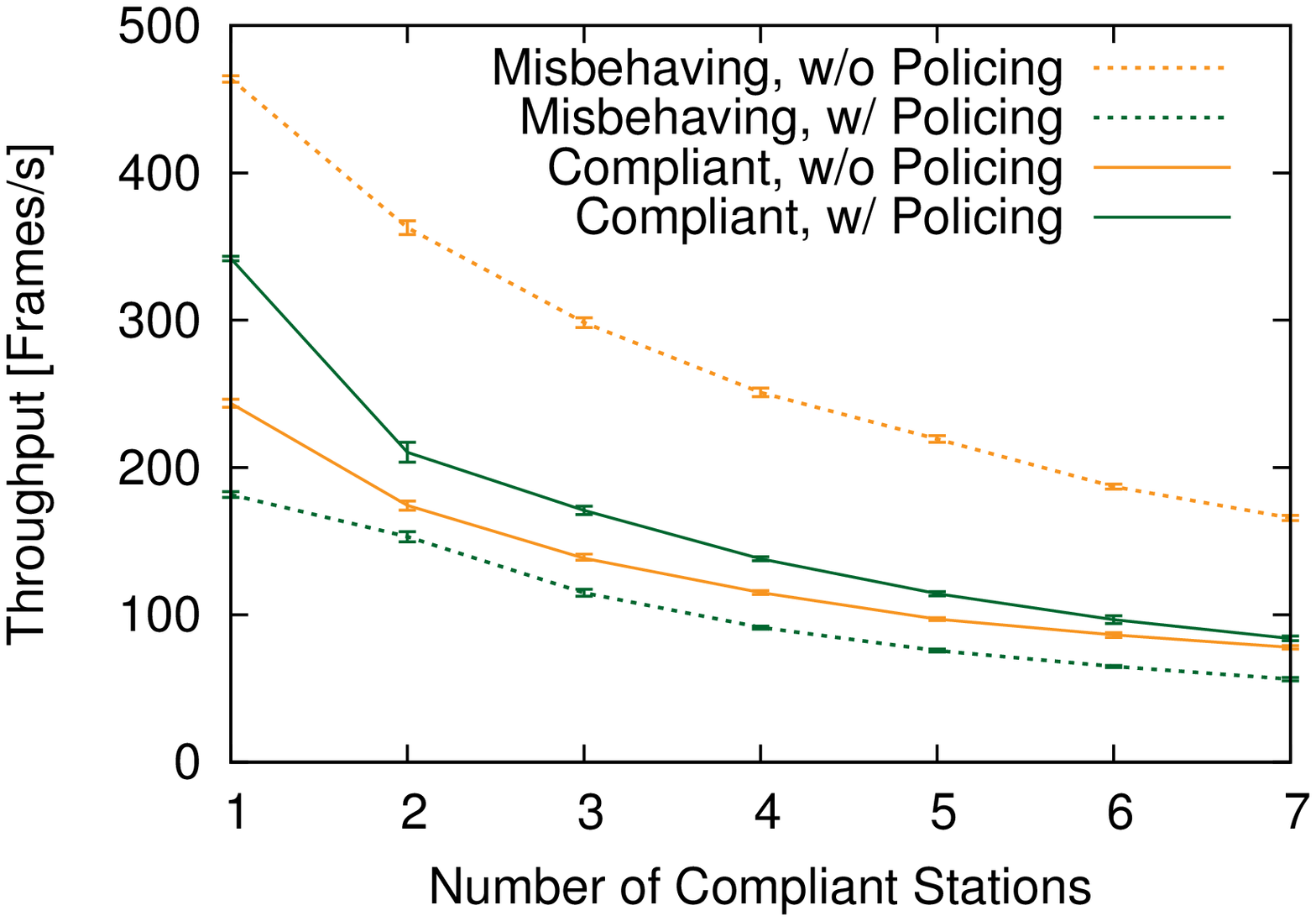}
\end{center}
\vspace*{-0.25em}
\caption{WLAN consisting of one misbehaving client with CW$_{\text{min}}$ half the default value and an increasing number of compliant stations. All clients always have a 1,000-byte packet to transmit at 11Mb/s (802.11b). Average and 95\% confidence intervals of the attempt rate (above) and throughput (below) attained by the misbehaving station and one fair user, when the AP operates with and without the proposed policing scheme. Experimental data.}
\label{fig:increasing-fair}
\vspace*{-0.5em}
\end{figure}

In Fig.~\ref{fig:increasing-fair} we show the attempt rate and throughput of the selfish station and that of one fair client, with a standard AP as well as with an AP executing our algorithm. Observe that the performance of the selfish user decreases as the network size increases, but is constantly significantly above that of a fair client if no action is taken to counteract the greedy behaviour. In contrast, when the AP runs our policing algorithm, the attempt rate of the misbehaving user never exceeds that of a fair client (observe the overlapping dark lines in the top sub-figure), while their throughput performance falls below that of fair clients in all circumstances.

We conclude that the network size does not impact the performance of our algorithm, which effectively penalises misbehaving clients even in denser topologies.

\begin{figure}[t]
\begin{center}
\includegraphics[height=0.73\columnwidth,angle=270]{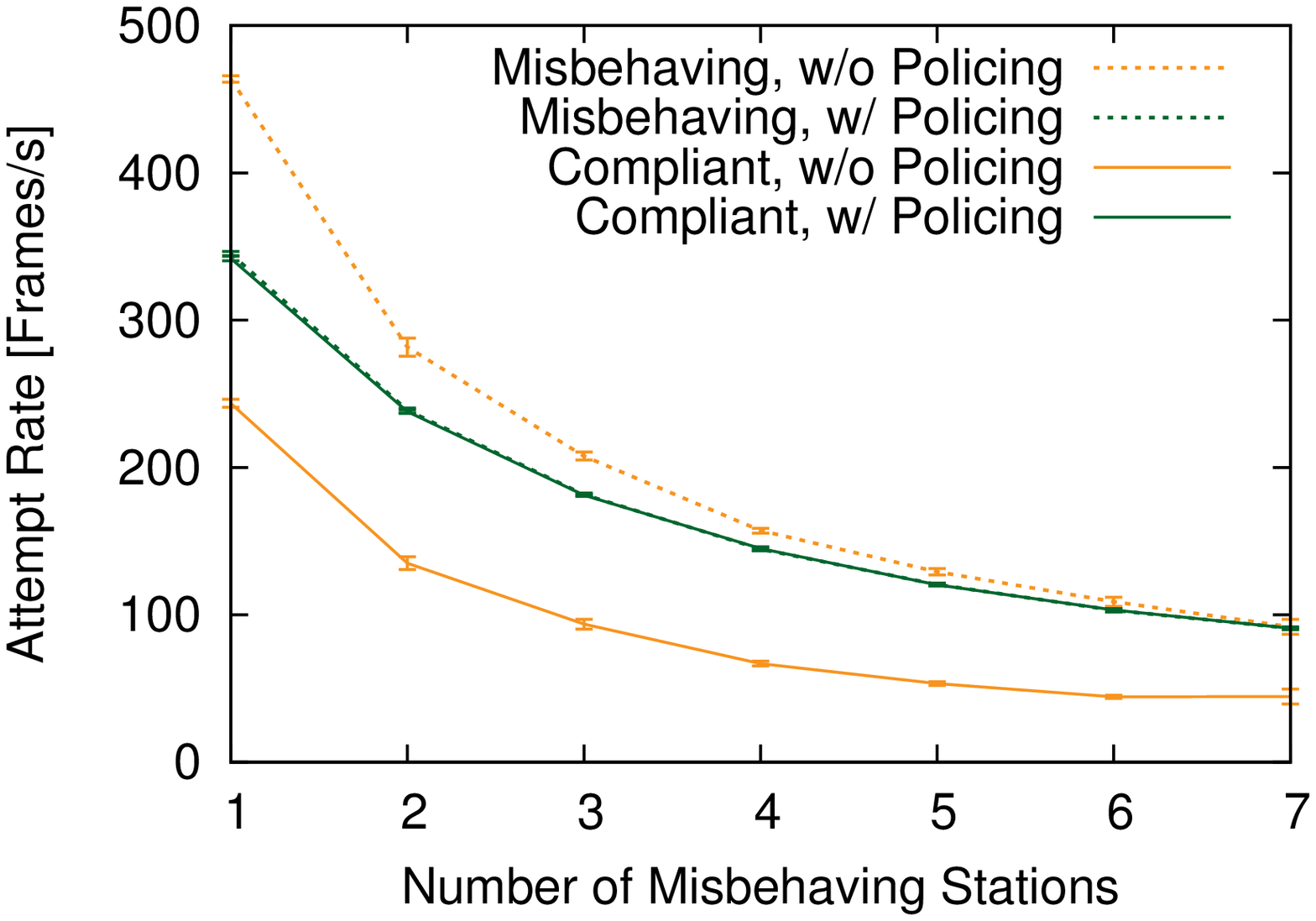}
\vspace*{-0.3em}
\includegraphics[height=0.73\columnwidth,angle=270]{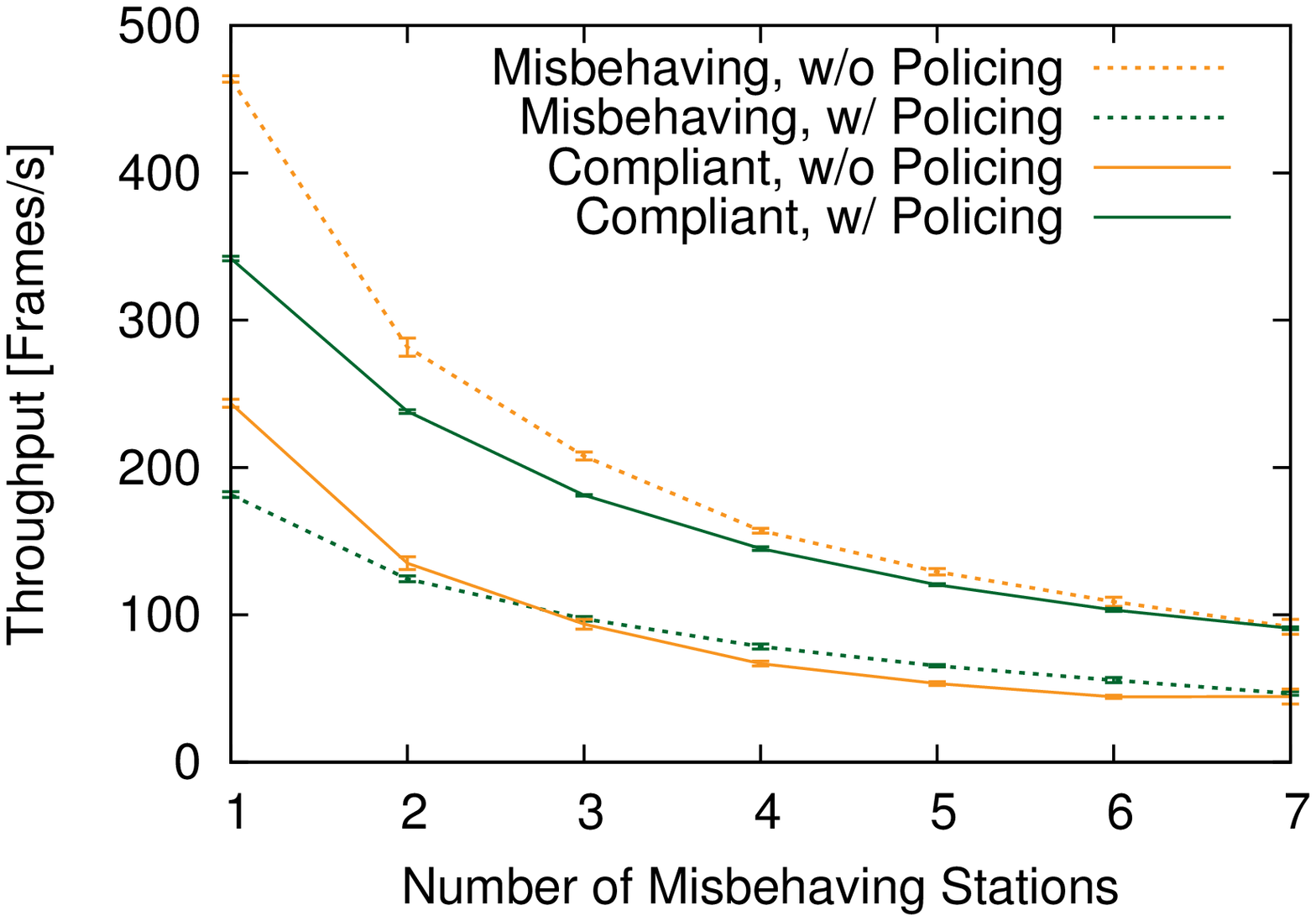}
\end{center}
\caption{WLAN consisting of one compliant station and an increasing number of misbehaving users with CW$_{\text{min}}$ half the default value. All stations are backlogged with 1,000-byte packets and transmit at 11Mb/s (802.11b). Average and 95\% confidence intervals of the attempt rate (above) and throughput (below) attained by the fair client and one selfish user, when AP operates with and without the proposed policing scheme. Experimental data.}
\label{fig:attackers}
\end{figure}

\begin{figure}[!t]
\begin{center}
\includegraphics[height=0.73\columnwidth,angle=270]{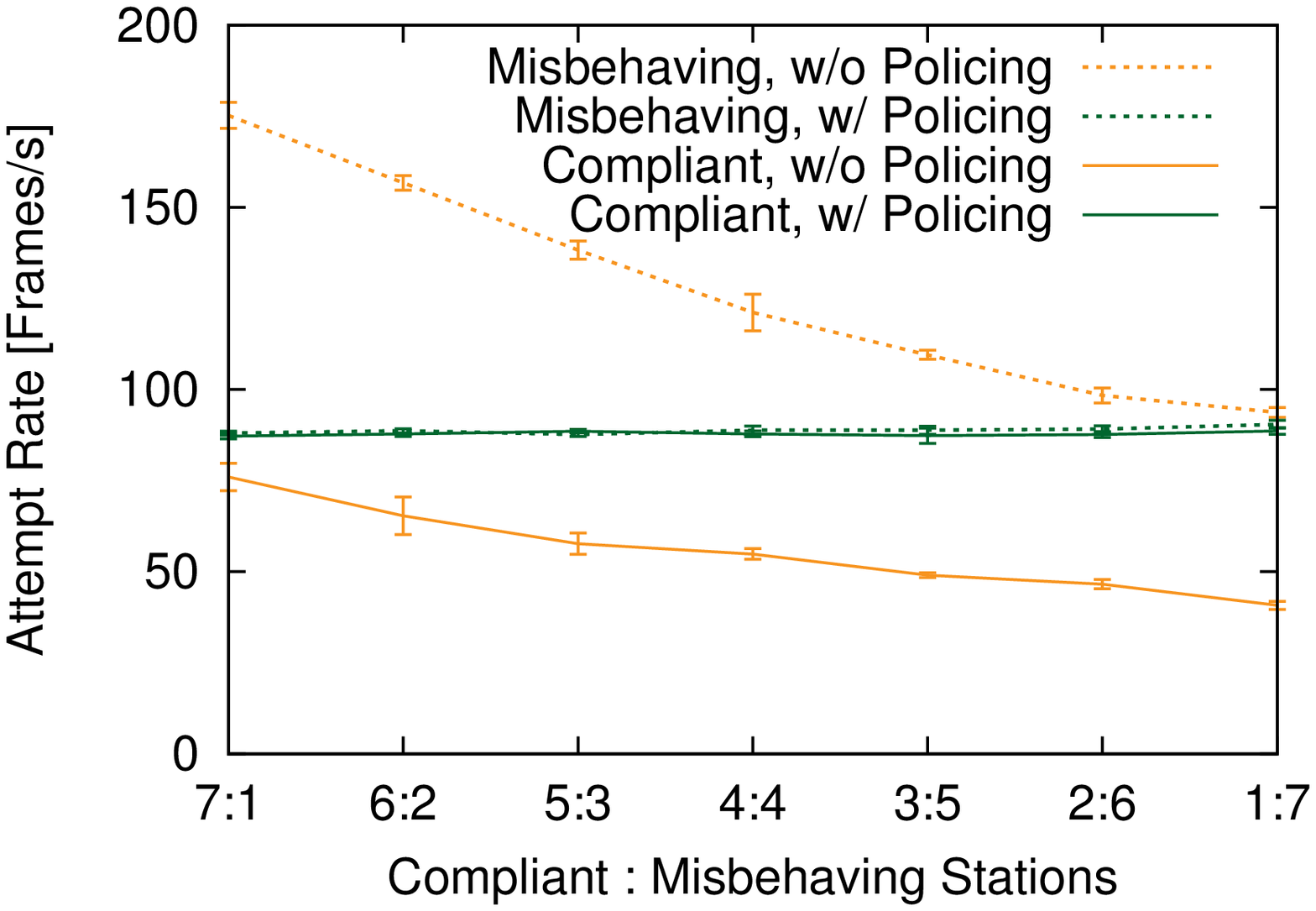}
\vspace*{-0.3em}
\includegraphics[height=0.73\columnwidth,angle=270]{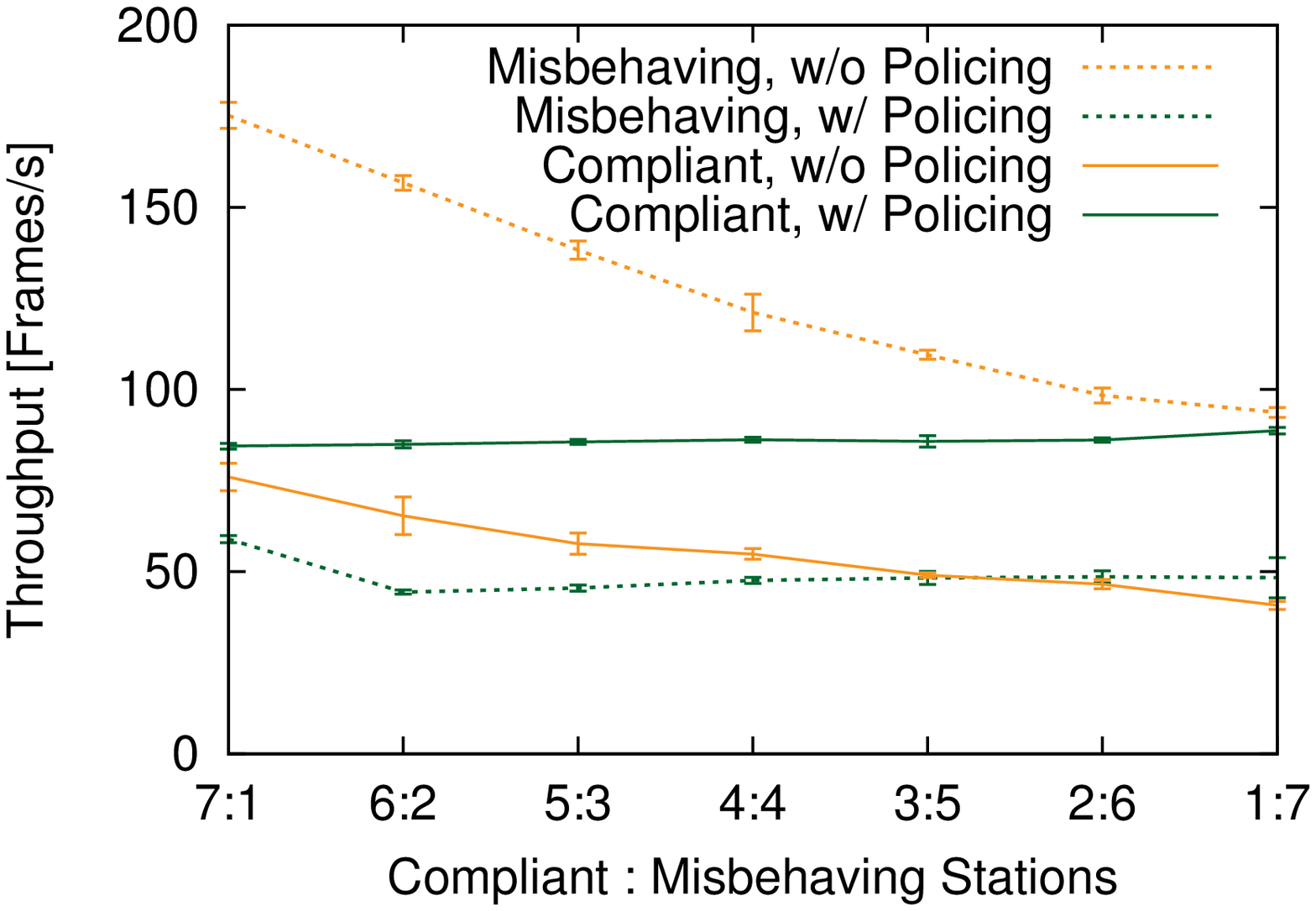}
\end{center}
\caption{WLAN consisting of eight backlogged clients transmitting 1,000-byte packets at 11Mb/s (802.11b). The ratio of compliant:misbehaving stations is varied. Selfish users contend with CW$_{\text{min}}$ half the default value. Average and 95\% confidence intervals of the attempt rate (above) and throughput (below) attained by a fair and a misbehaving station, when AP operates with and without the proposed policing scheme. Experimental data.}
\label{fig:mixed}
\vspace*{-1.5em}
\end{figure}

\vspace*{-0.25em}
\subsection{Multiple Misbehaving Clients}
In what follows, we study the performance of the proposed policing algorithm when multiple misbehaving clients are present in the WLAN. Here, we aim to understand whether the presence of a large number of selfish users could influence the penalty update of our algorithm. We demonstrate that, despite its prevalence, such behaviour will not be regarded as fair by the policing scheme. We use the same methodology as in the previous subsection, running 3-minute tests for each network scenario and conducting 10 independent experiments for each case. We measure the average performance of both fair and misbehaving stations in terms of attempt rate and throughput. 

First let us consider the case where only one station is fair and increase the number of selfish clients present in the network. The results of these experiments are depicted in Fig.~\ref{fig:attackers}, where we plot the attempt rate and throughput of the fair station and that of one non-compliant station, with and without the policing algorithm running at the AP. We observe that also in these scenarios, the policing algorithm equalises the attempt rate of all stations while the throughput performance of non-compliant users is effectively reduced.

In addition, we examine a network with a fixed number of clients ($n=8$) and vary the proportion of fair/misbehaving stations. The attempt rate and throughput of one client within each category is shown in Fig.~\ref{fig:mixed} when the AP operates with and without the proposed policing scheme. The obtained results further confirm the effectiveness of our approach in the presence of several misbehaving stations.

\revs{\subsection{Dynamic Network Conditions}}
\revs{
We consider next a scenario with network dynamics where fair and misbehaving clients join and leave the WLAN at different times. Our goal here is twofold: \emph{(i)} we verify that our proposal adapts quickly to changes in the network topology, and \emph{(ii)} we demonstrate the algorithm carries forward the penalty of selfish users when those leave the network. To this end, we conduct an experiment with the AP running our policing scheme and four backlogged client stations, as follows. Two fair stations connect to the WLAN and start transmitting to the AP at $t=0$s. After 100s, a misbehaving station (S3) joins the network, contending with a CW$_\text{min}$ parameter half the default value. At $t=200s$ another standard compliant station (S4) connects to the WLAN. Finally, S3 leaves the network after transmitting for 200s and S4 disassociates 100s later.
}

\begin{figure}[t]
\begin{center}
\includegraphics[height=0.79\columnwidth,angle=270]{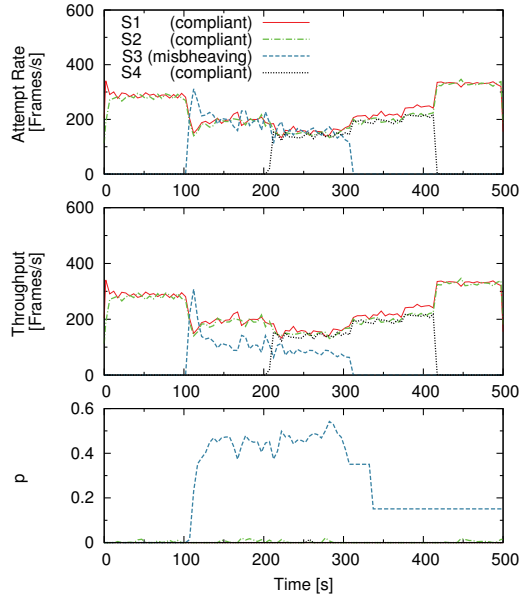}
\end{center}
\caption{\revs{WLAN with dynamic topology: two compliant stations are joined by a misbehaving one (CW$_{\text{min}}$ half the default value) and subsequently by a third fair client. Stations S3 and S4 transmit for 200s each and then leave the network. The AP runs the proposed policing scheme. Time evolution of the attempt rate (above), throughput (middle) and penalty applied by the proposed policing algorithm (below) for each client. Experimental data.}}
\label{fig:dynamic}
\end{figure}

\revs{
The result of this experiment is depicted in Fig.~\ref{fig:dynamic} where we plot the time evolution of the attempt rate, throughput and penalty corresponding to each client. We can see clearly that our algorithm quickly detects and starts penalising the misbehaving station, equalising the attempt rates in a few iterations. As the forth client joins, our solution re-estimates the maximum achievable attempt rate and continues penalising the selfish user, without affecting the performance of the new station. Lastly, as the cheater leaves the network, the penalty is preserved and carried forward to be applied when this client will reconnect. Thus we confirm that the performance of our algorithm is not affected by network dynamics and penalties are successfully carried forward.
}

\begin{figure*}[!t]
\normalsize
\centering
\includegraphics[width=0.43\columnwidth,angle=270]{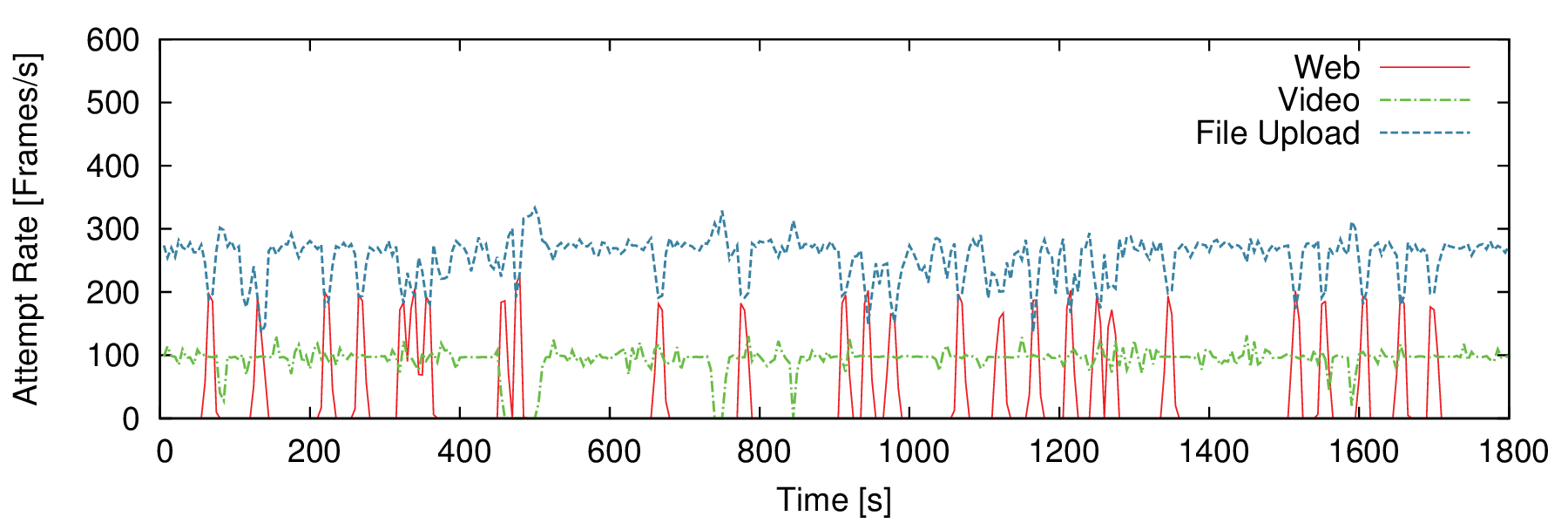}
\includegraphics[width=0.43\columnwidth,angle=270]{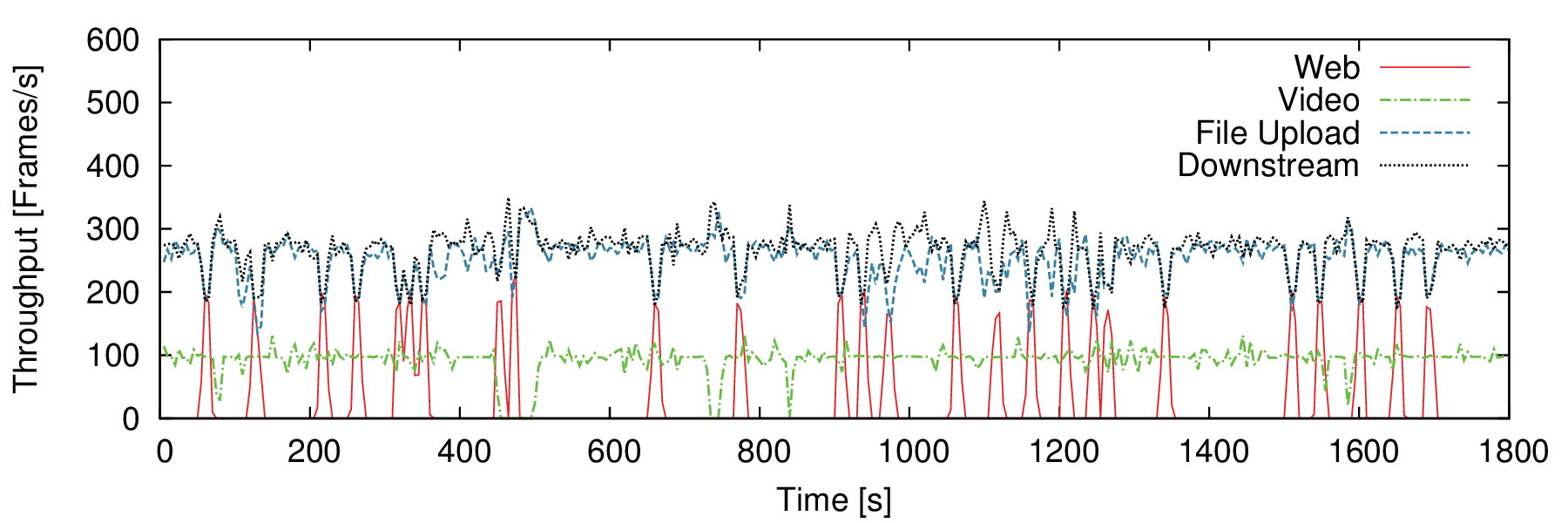}
\includegraphics[width=0.43\columnwidth,angle=270]{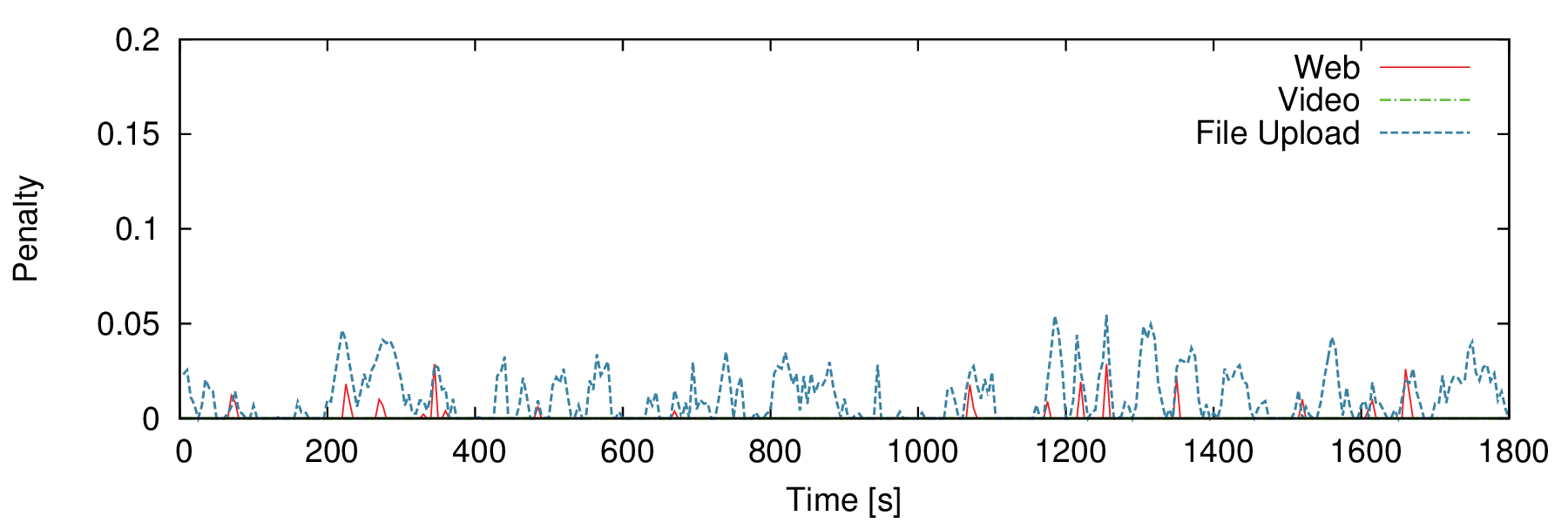}
\caption{WLAN consisting of four standard compliant stations generating heterogeneous traffic: file upload, web browsing, video streaming, system update (download). AP runs the proposed policing scheme. 30-minute snapshot of the attempt rate (above) and throughput (middle) attained by each flow, as well as the penalties applied by our algorithm (below). Experimental data.}
\label{fig:realtraf}

\end{figure*}

\subsection{Real Traffic}
Next, we investigate the performance of the policing algorithm in a more realistic scenario with heterogeneous traffic. We will show that the policing algorithm does not unnecessarily penalise fair clients that have increased demands and attain higher transmission rates simply due to the reduced activity of the other contenders. 

Towards this end, we consider a network with $n=4$ clients, the first one uploading a large file, the second generating web traffic, the third streaming a video file and the last performing a system update. 
To emulate the file upload, we generate saturated traffic using \texttt{iperf} on the first client. The second station establishes finite size TCP connections, alternating between periods of activity, during which a 2-Mbyte file is transferred, and silent periods exponentially distributed with mean $\lambda^{-1} = 60$s \cite{barford98}. The third station streams a MPEG-4 encoded version of ``Resident Evil: Apocalypse'' at 1 Mb/s using the VLC media player \cite{vlc}. To emulate the activity of the forth station, we use a backlogged \texttt{iperf} downstream session from the AP to the client. In this scenario, as the AP is always fair, we use the downstream flow to estimate the fair throughput. We run the experiment for a total duration of 1~hour, measuring for each flow the attempt rate, throughput and penalty applied by our policing scheme.

In Fig.~\ref{fig:realtraf} we plot a 30-minute snapshot of the network operation in this experiment, showing the time evolution of the aforementioned performance metrics for each client station. First, we observe that the penalty stays at zero most of the time for all stations, only with infrequent and small variations (below 0.05) above zero. Second, the medium-quality video flow sees its bandwidth demand satisfied most of the time. Third, the bandwidth demanding upload and download flows equally share the remaining available air time. Lastly, the spurious web traffic experiences similar performance to that of the other data flows whenever they are competing.

We conclude that indeed the proposed policing algorithm does not penalise stations that generate more traffic than their competitors as long as they comply with the MAC configuration defined by the 802.11 standard.

\begin{figure}[t]
\begin{center}
\includegraphics[height=0.70\columnwidth,angle=270]{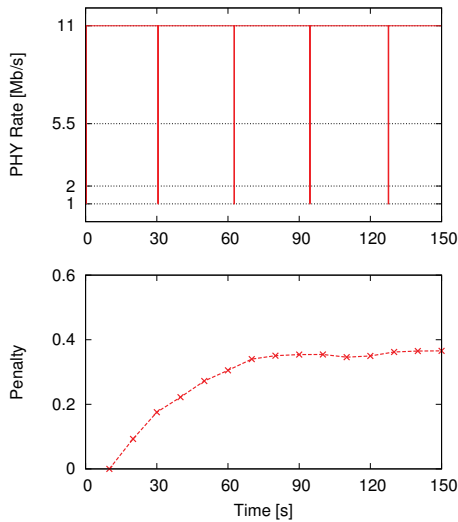}
\end{center}
\caption{WLAN consisting of three saturated stations that transmit 1,000-byte packets using the IEEE 802.11 HR/DSSS physical layer. Two stations are compliant and transmit at 11Mb/s, the third is misbehaving (CW$_\text{min}$ halved) and runs the Minstrel RC algorithm. \revs{Clients can choose from the following set of PHY bit rates for transmission: \{1, 2, 5.5 and 11\} Mb/s.} The proposed policing scheme is executed at the AP. PHY rates selected by the selfish client (above) and penalty applied (below) over a 150s period. Experimental data.}
\label{fig:rca}
\end{figure}

\section{Non-ideal Channel Effects}
\label{sec:nonideal}

We also investigate the performance of our implementation under several challenging situations that occur frequently in practice. Specifically, we verify that the proposed algorithm has no negative impact on rate switching decisions taken by state-of-the-art rate control algorithms and demonstrate the potential of our scheme to alleviate unfairness issues that arise due to the PHY/MAC interactions occurring in the presence of the capture effect.

\subsection{Rate Control}
We study the behaviour of a rate control algorithm executed at a greedy client that manipulates their MAC configuration and is being penalised by our policing algorithm to counteract their misbehaviour. Our goal here is to verify that rate control (RC) algorithms will not wrongly interpret suppressed ACKs as losses caused by poor channel conditions and thus will not trigger downgrades of the PHY rate. This is particularly important, since unnecessarily selecting a lower modulation scheme can be wasteful of channel time and have a significant impact on the overall network utility \cite{heusse03}. 

To this end, we consider again a simple scenario with two fair clients and one misbehaving station that uses a CW$_\text{min}$ parameter half the standard recommended value. In this experiment, the selfish client runs the Minstrel rate control algorithm, \revs{which is the default mechanism implemented by \texttt{mac80211} drivers on Linux systems since kernel version 2.6.29 (March 2009 to date)}, and the AP executes the proposed policing scheme. \revs{Note that Minstrel \cite{minstrel}, SampleRate \cite{bicket05} and other commonly used rate control schemes work by sampling the mean transmission time at different PHY rates. Since our ACK dropping scheme impacts on all PHY rates in the same way, it will inflate the transmission times for all rates in the same way, and consequently we expect the rate control scheme will still pick the rate with shortest transmission time. Similarly, schemes that make decisions based on SNR or related indicators will not be mislead by ACK dropping \cite{hrca}.}

We examine the time evolution of the penalty applied by our algorithm to the cheater, as well as the rate selected by Minstrel during the operation of our scheme. As shown in Fig.~\ref{fig:rca}, increasing the penalty does not influence the rate selection decisions taken by the rate control algorithm, since packets are transmitted almost always at the maximum rate (11~Mb/s) and lower rates are only periodically sampled (approx. every 30s), with only a couple of frames. 

\begin{figure}[t]
\begin{center}
\includegraphics[height=0.72\columnwidth,angle=270]{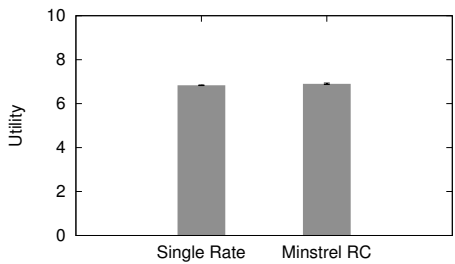}
\end{center}
\caption{WLAN consisting of three saturated stations that transmit 1,000-byte packets using the IEEE 802.11 HR/DSSS physical layer. Two stations are compliant and transmit at 11Mb/s and the third is misbehaving (CW$_\text{min}$ halved). The proposed policing scheme is executed at the AP. Network utility when the misbehaving client runs the Minstrel RC algorithm and uses a single PHY rate for transmission respectively. Experimental data.}
\label{fig:utility}
\vspace*{-1em}
\end{figure}

To verify that indeed the network utility is not affected when policing is applied to selfish stations, we also plot in Fig.~\ref{fig:utility} this metric for the same experiment, as well as for the case when the misbehaving client does not perform rate adaptation and all stations transmit at a single rate, e.g. 11Mb/s . Note that we compute the network utility as in \cite{kelly97}, i.e. the sum of the natural logarithm of the individual throughputs, which is considered a good measure of proportional fairness \cite{checco11}. From the results in Fig.~\ref{fig:utility} we conclude that our policing algorithm does not tamper with the operation of current rate control mechanisms and thus has no negative impact on the network utility when penalties are applied to non-compliant client stations.

\begin{figure}[t]
\vspace*{-0.6em}
\begin{center}
\includegraphics[height=0.72\columnwidth,angle=270]{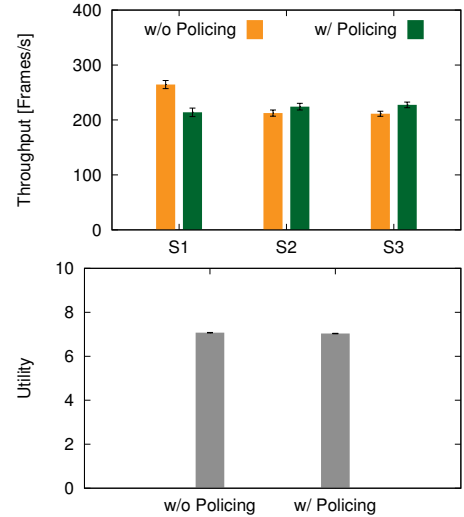}
\end{center}
\caption{WLAN consisting of three compliant stations always having 1,000-byte packets to transmit using the IEEE 802.11 HR/DSSS physical layer at 11Mb/s. Station (S1) is located next to the AP. Stations S2 and S3 are placed at a distance four times longer, thus S1 can capture the channel over S2 and S3. Average and 95\% confidence interval of per-station throughput shown above with a regular AP (light bars) and an AP running the proposed policing scheme (dark bars). Network utility shown below, with and without policing. Experimental data.}
\label{fig:capture}
\vspace*{-1em}
\end{figure}

\subsection{Capture Effect}

We investigate a scenario where all stations obey the standard specification, but experience different performance due to their placement relative to the AP. Specifically, we are interested in checking whether our policing scheme can improve fairness when a client that is located closer to the AP captures the channel while transmitting simultaneously with stations that reside farther away. This effect is frequently encountered in practical deployments and can cause significant unfairness, as already documented in e.g. \cite{Patras:2012:WoWMoM,Patras:2013:PMC}. 

For this purpose, we examine again the performance of a network with three fair stations, but this time with one station (S1) located next to the AP and the other two (S2 and S3) at similar, but four times longer distances. In the top plot of Fig.~\ref{fig:capture} we show the average throughput attained by each client in this scenario, with and without our policing algorithm running at the AP. Observe that without policing S1 achieves significantly better performance than the other two clients with a standard AP (light bars). On the other hand, when the AP executes our policing algorithm, the attempt rate of the station positioned near the AP will be reduced and consequently all stations will attain nearly identical throughputs (dark bars). Note that this correction of the throughput distribution among clients comes at no network utility cost, as we show in the lower plot of Fig.~\ref{fig:capture}.

We conclude that our policing algorithm does not only combat MAC misbehaviour, but can also mitigate unfairness that arises in real deployments due to the PHY/MAC interactions.

\section{Conclusions}
\label{sec:conclusions}
In this paper we introduced a policing scheme that penalises MAC misbehaviour and preserves fairness in wireless networks. The proposed algorithm is executed at the AP and does not require any modification to compliant devices. We established the convergence of our algorithm, as well as its robustness to sophisticated misbehaviour strategies that seek to game its operation. We presented a practical implementation on off-the-shelf hardware and demonstrated the effectiveness of our proposal by conducting extensive experiments in a real wireless LAN, over a wide range of network conditions and misbehaviour scenarios. The results obtained show that our policing algorithm drives selfish users into compliant operation, regardless of the type of misbehaviour employed, and does not penalise compliant clients that consume more air time than lightly loaded stations. In addition to that, we showed that our solution has no negative impact on current rate control algorithms and can alleviate unfairness incurred by the 
physical layer capture effect.

\section*{Acknowledgements}
The authors wish to thank Francesco Gringoli for his valuable support with OpenFWWF and Ken Duffy for his thoughtful comments that helped improving this article.

\bibliographystyle{IEEEtran}
\bibliography{references}

\end{document}